\documentclass[twocolumn,nodate,showpacs,preprintnumbers,nofootinbib,amsmath,amssymb,aps,prd]{revtex4-1}

\usepackage{graphicx}
\usepackage{amsmath,amssymb}

\usepackage[english]{babel}
\usepackage{graphics}
\usepackage{color}
\usepackage{wrapfig}
\usepackage{epsfig}
\usepackage{pst-grad} 
\usepackage{times}

\def\f{\frac}
\def\be{\begin{equation}}
\def\ee{\end{equation}}
\def\ba{\begin{eqnarray}}
\def\ea{\end{eqnarray}}

\def\s1pr{\mathcal{I}^{1+}_{\rm R}}

\def\spr{\mathcal{I}^{+}_{\rm R}}

\def\sml{\mathcal{I}^{-}_{\rm L}}
\def\smr{\mathcal{I}^{-}_{\rm R}}
\def\sprb{\bar{\mathcal{I}}^+_{\rm R}}
\def\mbt{M_{\rm Bondi}^{\rm Trad}}
\def\mbatv{M_{\rm Bondi}^{\rm ATV}}
\def\ADM{{\rm ADM}}

\def\Bondi{{\rm Bondi}}
\def\fatv{F^{\rm ATV}}

\def\GDH{{\rm GDH}}
\def\N{\bar{N}}

\def\h{\hat}

\def\Dp{\partial_{+}}
\def\Dm{\partial_{-}}
\def\dd{{\rm d}}

\def\a{\mathbf{a}}

\def\k{\kappa}

\newcommand{\secintro}{I}
\newcommand{\secmodel}{II}
\newcommand{\secscaling}{III}
\newcommand{\secresults}{IV}
\newcommand{\secconclusions}{V}

\begin{document}

\title{Surprises in the Evaporation of 2-dimensional Black Holes}

\author{Abhay Ashtekar${}^{1}$ }
\author{Frans Pretorius ${}^{2}$}
\author{Fethi M. Ramazano\u{g}lu${}^{2}$}
\affiliation{${}^1$\! Institute for Gravitation and the Cosmos \&
     Physics Department, Penn State, University Park, PA 16802,
    USA\\
${}^2$\! Department of Physics, Princeton University, 08544,
Princeton, NJ, USA }	

\begin{abstract}

Quantum evaporation of Callan-Giddings-Harvey-Strominger (CGHS)
black holes is analyzed in the mean field approximation. This
semi-classical theory incorporates back reaction. Detailed
analytical and numerical calculations show that, while some of the
assumptions underlying the standard evaporation paradigm are borne
out, several are not. Furthermore, if the black hole is initially
macroscopic, the evaporation process exhibits remarkable universal
properties (which are distinct from the features observed in the
simplified, exactly soluble models). Although the literature on CGHS
black holes is quite rich, these features had escaped previous
analyses, in part because of lack of required numerical precision,
and in part because certain properties and symmetries of the model
were not fully recognized. Finally, our results provide support for
the full quantum gravity scenario recently developed by Ashtekar,
Taveras and Varadarajan.

\end{abstract}


\maketitle

\noindent{\bf{\em \secintro. Introduction.}} Since the early
nineties, a number of 2-dimensional (2D) black hole models have
been studied to gain further insight into the quantum dynamics
of black hole evaporation. Physically, the most interesting
among them is due to Callan-Giddings-Harvey-Strominger (CGHS)
\cite{cghs}. Simplified versions of this model are exactly
soluble but also have important limitations discussed, e.g., in
\cite{reviews,hs}. Therefore results obtained in those models
are not reliable indicators of what happens in the full CGHS
dynamics. In this letter we present key results from a new
analysis of CGHS black holes using a mean-field or
semi-classical approximation. These findings are surprising in
two respects. First, several features of the standard CGHS
paradigm ~\cite{reviews} of quantum evaporation are not
realized. Second, black holes resulting from a prompt collapse
of a large Arnowitt-Deser-Misner (ADM) mass exhibit rather
remarkable behavior: after an initial transient phase, dynamics
of various physically interesting quantities at right future
null infinity $\spr$ flow to {\em universal curves},
independent of the details of the initial collapsing matter
distribution. This universality strongly suggests that
information in the collapsing matter on $\smr$ can {\em not} in
general be recovered at $\spr$. However, we also find strong
evidence supporting the scenario of \cite{atv} in which the
$S$-matrix from (left past infinity) $\sml$ to $\spr$ is
unitary. This distinction between unitarity and information
recovery
is a peculiarity of 2D.

In this letter we summarize the main results. An extensive
treatment can be found in ~\cite{apr}; details about the
numerics in~\cite{pr}; and a more thorough investigation of the
full quantum issues in~\cite{vat}.
\smallskip

\noindent{\bf{\em \secmodel. Model.}} In the CGHS model, geometry is
encoded in a physical metric $g$ and a dilaton field $\phi$, and
coupled to $N$ massless scalar fields $f_i$. Since we are in 2D with
$\mathbb{R}^2$ topology, we can fix a fiducial flat metric $\eta$
and write $g$ as $g^{ab} = \Omega \eta^{ab}$. Then it is convenient
to describe geometry through $\Phi := e^{-2\phi}$ and $\Theta :=
\Omega^{-1} \Phi$. The model has 2 constants, $\kappa$ with
dimensions $[L]^{-1}$ and $G$ with dimensions $[ML]^{-1}$.

Our investigation is carried out within the mean field approximation
(MFA) of \cite{atv,vat} in which one ignores quantum fluctuations of
geometry but not of matter. To ensure a sufficiently large domain of
validity, we must have large $N$ and we assume that each scalar
field $f_i$ has the same profile on $\smr$. Black hole formation and
evaporation is described entirely in terms of non-linear partial
differential equations.
\begin{figure}
\begin{center}
\includegraphics[width=2.00in,angle=0]{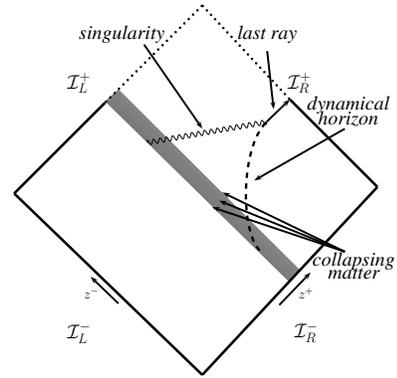}
\caption{A Penrose diagram of an evaporating CGHS black hole in the
mean field approximation (MFA). The incoming state is the vacuum on
$\sml$, and left moving  matter distribution on $\smr$. The
collapse creates a generalized dynamical horizon (GDH), which
subsequently evaporates. Quantum radiation fills the spacetime
to the causal future of matter. Inside the GDH, a singularity forms
in the geometry. It meets the GDH when the latter shrinks to zero
area. The ``last ray'' emanating from this meeting point is a future
Cauchy horizon.}
\vskip-0.4cm
 \label{fethi-sketch}
\end{center}
\end{figure}
Denote by $z^\pm$ the advanced and retarded null coordinates of
$\eta$ so that $\eta_{ab} = 2 \partial_{(a}z^+\,
\partial_{b)}z^-$. We will set $\partial_{\pm} \equiv
\partial/\partial z^\pm$. Then we have the evolution equations
\be \label{de1}\Box_{(\eta)}\, f_i = 0\quad \Leftrightarrow
\quad\Box_{(g)}f_i = 0.\ee
for matter fields, and
\ba \label{de2}\Dp\,\Dm\, \Phi + \k^2 \Theta = G\,
\langle\h{T}_{+-}\rangle \equiv {\N G\hbar}\, \Dp\, \Dm\, {\ln
(\Phi\Theta^{-1})}\nonumber \\
\label{de3} \Phi \Dp\,\Dm {\ln \Theta} =  - G\,
\langle\h{T}_{+-}\rangle \equiv -{\N G\hbar}\, \Dp\, \Dm\, {\ln
(\Phi\Theta^{-1})} \ea
for geometric fields $\Theta, \Phi$. The terms on the right side are
quantum corrections to the classical equations due to conformal
anomaly and encode the back reaction of quantum radiation. As in 4D
general relativity there are constraints which are preserved by the
evolution equations:
\ba \label{ce1}  -\Dm^2\, \Phi + \Dm\,\Phi \Dm\, \ln \Theta \,&=&\,
G\,\langle \h{T}_{- -}\rangle\nonumber\\
-\Dp^2\, \Phi + \Dp\,\Phi \Dp\, \ln \Theta \,&=&\, G\, \langle
\h{T}_{++}\rangle\, . \ea
Here, $\N:=N/24$ and $\langle\h{T}_{ab}\rangle$ denotes the
expectation value of the stress-energy tensor of the $N$ fields
$f_i$.

We solve this system of equations as follows. As is standard in the
CGHS literature, we assume that prior to $z^+\!=\!0$ the space-time
is given by the classical vacuum solution and matter falls in from
$\smr$ after that (see Fig.~\ref{fethi-sketch}). Therefore, to
specify consistent initial data, it suffices to choose a matter
profile
$f_+(z^+)$ on $\smr$ , and solve for the initial $(\Theta,\Phi)$
using~(\ref{ce1}). We then evolve $(\Theta,\Phi)$ to the future of
the initial data surfaces using~(\ref{de2}). Trivially,
$f_i(z^+,z^-)=f_+(z^+)$ from (\ref{de1}).

We now discuss the interpretation of solutions to these equations
via horizons, singularities and the Bondi mass. Note first that in
analogous 4-dimensional (4D) spherically symmetric reductions,
$\Phi$ is related to the radius $r$ by $\Phi = \k^2 r^2$
\cite{reviews,apr}. Therefore, a point in the CGHS space-time
$(M,g)$ is said to be \emph{future marginally trapped} if $\Dp \Phi$
vanishes and $\Dm \Phi$ is negative there \cite{reviews,sh}. The
quantum corrected ``area'' of a trapped point is given by $\a :=
(\Phi - 2\N G\hbar)$. 
The world-line of these marginally trapped points forms a
\emph{generalized dynamical horizon} (GDH). As time evolves, this
area \emph{shrinks} because of quantum radiation (hence
`generalized': the world-line is time-like rather than space-like).
The area finally shrinks to zero. The MFA equations are formally
singular where $\Phi=2\N G\hbar$; thus at the end-point of
evaporation the GDH meets a space-like singularity.  The `last ray'
---the null geodesic from this point to $\spr$--- is the future
Cauchy horizon of the semi-classical space-time. See Fig.
\ref{fethi-sketch}.

We assume (and this is borne out by the simulations) that the
semi-classical space-time is asymptotically flat at $\spr$ in the
sense that, as  $z^+ \rightarrow \infty$, the field $\Phi$ has the
following behavior along $z^-$ = const lines
\be  {\Phi} = A(z^-)\, e^{\k z^+} + B(z^-) +
O(e^{-\k z^+}),
\label{asym} \ee
where $A$ and $B$ are smooth functions of $z^-$. A similar expansion
holds for $\Theta$. The physical semi-classical metric ${g}_{ab}$
admits an \emph{asymptotic} time translation $t^a$. Its affine
parameter $y^-$ is given by $ e^{-\k y^-} =\, A(z^-)$. Up to an
additive constant, $y^-$ serves as the unique physical time
parameter at $\spr$. The MFA equations imply that there is a balance
law at $\spr$ \cite{atv,vat}, motivating new definitions of a Bondi
mass $\mbatv$ and a manifestly positive energy flux $\fatv$:
\ba
\label{mass}\mbatv &=& \f{\dd{B}}{\dd y^-} + \k {B}\, +\, {\N
\hbar G}\, \big(\f{\dd^2 y^-}{\dd z^{-2}}\, (\f{\dd y^-}{\dd
z^-})^{-2}\, \big)\, \\
\label{flux}  \fatv &=&  \f{\N \hbar G}{2}\, \big[\f{\dd^2 y^-}{\dd
z^{-2}}\, (\f{\dd y^-}{\dd z^-})^{-2}\,\,\big]^2\, , \ea
so that $\dd(\mbatv)/\dd y^-= -\fatv$.
In the classical theory ($\hbar =0$), there is no energy flux at
$\spr$, and $\mbatv$ reduces to the standard Bondi mass formula,
which includes only the first two terms in~(\ref{mass}). Previous
literature \cite{cghs,reviews,sh,st,dl} on the CGHS model used this
classical expression also in the semi-classical theory. But we will
see that this traditionally used Bondi mass, $\mbt$, is physically
unsatisfactory.
\smallskip

\noindent{\bf{\em \secscaling. Scaling and the Planck Regime.}} It
turns out that the mean field theory admits a scaling symmetry. To
express it explicitly, let us regard all fields as functions of
$z^\pm$. Then, given any solution $(\Theta, \Phi, N, f_+)$ to all
the field equations and a positive number $\lambda$, $(\lambda
\Theta, \lambda \Phi, \lambda N, f_+)$ is also a solution (once
$z^-$ is shifted to $z^- + (\ln\lambda)/\k$) \cite{Ori:2010nn,apr}.
Under this transformation, we have
%
%
\be {g}^{ab}  \rightarrow {g}^{ab}, \,\,\, (M,\fatv,\a_{\GDH})
\rightarrow \lambda (M,\fatv,\a_{\GDH})\nonumber\ee
where $\a_{\GDH}$ denotes the area of the GDH, and $M$ is either the
Bondi mass $\mbatv$ or the ADM mass $M_{\ADM}$. This symmetry
implies that, \emph{as far as space-time geometry and energetics are
concerned, only the ratios $M/N$ matter.} Thus, whether a black hole
is `macroscopic' or `Planck size' depends on the ratios $M/N$ and
$\a_{\GDH}/N$ rather than on the values of $M$ or $\a_{\GDH}$
themselves. Hence we are led to define
\ba (M^\star,M^\star_{\Bondi},F^\star)&=&(M_{\ADM},\mbatv,\fatv)/{\N}, \ \
\nonumber\\
{\rm and}\,\,\,\, m^\star &=& M^\star_{\Bondi}|_{\hbox{\rm last
ray}}\, . \label{star}\ea
To compare these quantities to the Planck scale we first note
that there are some subtleties because $G\hbar$ is
dimensionless in 2D and careful considerations lead us to set
$M_{\rm Pl}^2 = \hbar\kappa^2/G$ and $\tau_{\rm Pl}^2 =
G\hbar/\kappa^2$ \cite{apr}. We can now regard a black hole as
macroscopic if its evaporation time is much larger than the
Planck time. Using the fact that, in the external field
approximation, the energy flux is given by $F_{\rm Haw} =
(\N\hbar\k^2/2)$, this condition leads us to say that \emph{a
black hole is macroscopic if $M^\star \gg G\hbar\,\, M_{\rm
Pl}$.} Note that the relevant quantity is $M^\star$ rather than
$M$. The precise nature of this scaling property was not
appreciated until recently. For example, in~\cite{ps} it was
noted that $N$ could be ``scaled out'' of the problem and that
the results are ``qualitatively independent of N'' whereas in
fact for a given $M$ they can vary significantly as $N$
changes. Similarly, the condition that a macroscopic black hole
should have large $M/N$ appears in \cite{st}. But it was
arrived at by physical considerations involving static
solutions rather than an exact scaling property of full
equations.
\smallskip

\noindent{\bf{\em \secresults. Results.}} Here we describe some
key results from numerical solution of the CGHS equations
(\ref{de1})-(\ref{de3}). We consider two families of initial
data, most conveniently described in a ``Kruskal-like''
coordinate $\k x^+ = e^{\k z^+}$. The first is a collapsing
shell used throughout the CGHS literature so far,
\be \label{shell} \left(\partial f_+/\partial x^+\right)^2 =
\f{M^\star}{12} \, \delta\left(x^+ - 1/\k\right), \ee
parameterized by $M^\star$. The other is a smooth
($f_+(x^+)$ is $\mathcal{C}^4$), two parameter
$(\tilde{M}^\star,w)$ profile defined by
\be \label{shell_w} \textstyle{\int_0^{x^+}}\! \dd \bar{x}^+\,
(\frac{\partial f_+}{\partial \bar{x}^+})^2\, =
\f{\tilde{M}^\star}{12} \left( 1 -e^{\left( \kappa x^+ -1
\right)^2/w^2}\right)^4 \theta(x^+-1/\kappa), \ee
where $\theta$ is the unit step function, $w$ characterizes the
width of the matter distribution, and $\tilde{M}^\star$ is related
to the ADM mass via $M^\star\approx \tilde{M}^\star (1+1.39 \ w)$.
Unraveling of the unforeseen behavior required high precision
numerics \cite{pr}, 
which is crucial in the macroscopic mass limit that is of primary
importance. Numerical solutions from both classes of initial data
were obtained for a range of masses $M^\star$ from $2^{-10}$ to
$16$, a range of widths from $w=0$ to $w=4$, and $\N$ varying from
$0.5$ to $1000$. Since we are primarily interested in black holes
which are initially macroscopic, here we will focus on $M^\star \ge
1$ and, since the computations did bear out the scaling behavior, on
the case $\N =1$. We set $\hbar\!=G\!=\!\kappa\!=\! 1$.

Our numerical simulations show that, as expected, the semi-classical
space-time is asymptotically flat at $\spr$ but, in contrast to the
classical theory, $\spr$ is incomplete, i.e., $y^-$ has a finite
value at the last ray.
However, dynamics also exhibits some surprising features which can
be summarized as follows.

\begin{figure}
\begin{center}
\includegraphics[scale=0.65]{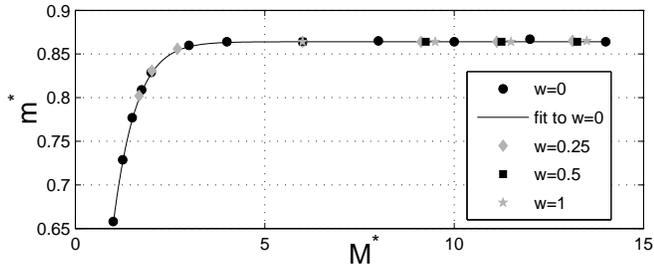}
\end{center}
\caption{The final mass $m^\star$ versus the
initial mass $M^\star$ (\ref{star}) for a variety
of initial data (\ref{shell}-\ref{shell_w}).
The curve fit to the data is $m^\star = \alpha\,
(1-e^{-\beta (M^\star)^\gamma})$, with $\alpha \approx 0.864$,
$\beta \approx 1.42$, and $\gamma\approx 1.15$.} \vskip-0.4cm
\label{mass_at_last_ray}
\end{figure}

First, the traditionally used Bondi mass $\mbt$ can become
\emph{negative and large even when the \GDH\ is macroscopic.}
For CGHS black holes, negative $\mbt$ was known to occur
\cite{tu} but only for black holes which are of Planck size
even before evaporation begins. 
For initially macroscopic black holes, the standard paradigm
assumed that $\mbt$ is positive and tends to zero as the \GDH\
shrinks (so that one can attach a `flat corner of Minkowski
space to the future of the last ray). Second, while the
improved Bondi mass, $\mbatv$, does remain positive throughout
evolution, at the last ray it can be large. In fact this `end
state' exhibits a universality shown in Fig
\ref{mass_at_last_ray} where $m^\star$, the final value of
$(M^\star_{\Bondi})$, is plotted against the rescaled ADM mass
$M^\star$ for a range of initial data. It is clear from the
plot that there is a qualitative difference between $M^\star
\gtrsim 4$ and $M^\star \lesssim 4$. In the first case the
value of the end point Bondi mass is universal, $m^\star
\approx 0.864$. For $M^\star <4$ on the other hand, the value
of $m^\star$ depends sensitively on $M^\star$. Thus in the MFA
it is natural to regard CGHS black holes with $M^\star \gtrsim
4$ as {\em macroscopic}, and those with $M^\star \lesssim 4$ as
{\em microscopic}. Numerical studies have been used in the past
to clarify properties of the CGHS model~\cite{dl,ps,tu,hs},
such as dynamics of the \GDH. However, they could not uncover
universal behavior because, in the present terminology, they
covered only \emph{microscopic} cases ($M^\star
\le 2.5$ in all prior studies). 
%
\begin{figure}
\begin{center}
\includegraphics[scale=0.65]{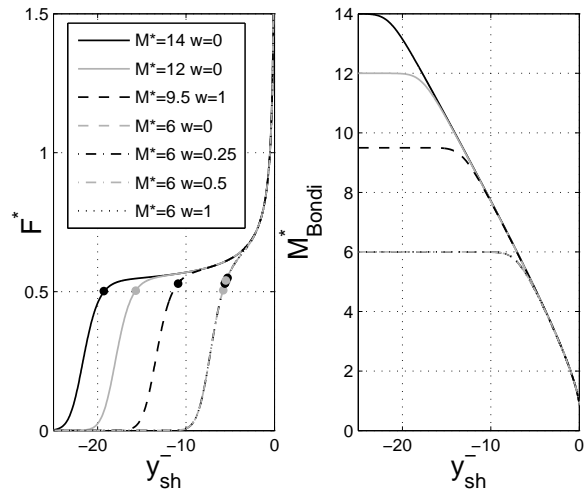}
\end{center}
\hskip-0.2cm
\caption{ $F^\star$ and $M^\star_{\Bondi}$ of Eq (\ref{star}) plotted
against $y^-_{\rm sh}$ for solutions with several values of
parameters $M^\star$ and $w$ of Eqs (\ref{shell})-(\ref{shell_w}).
In all cases $F^\star$ starts at $0$ in the distant past ($\kappa y^-_{\rm
sh} \ll -1$), and then joins a universal curve at a time that
depends on the initial mass. The time when the dynamical horizon
first forms is marked on each flux curve (which is later for larger
$w$, though note that the mass and flux curves for all the
$M^\star=6$ cases are indistinguishable in the figure).
We have not yet found an extrapolation of the flux to the last ray
$y^-_{\rm sh}=0$ that conclusively answers whether it is finite there.
However, all functions we tried that fit the data well have
a {\em finite integrated} flux. Moreover, when the flux starts rising
rapidly, we are still well within the regime where the numerical
solution converges, and we can follow the solution clearly into a
regime where the mass has reached its final value of
$m^\star\approx0.864$.}
\label{mass_flux}
\end{figure}

Third, for macroscopic ($M^\star \gtrsim 4$) black holes that form
{\em promptly}, after some early transient behavior, dynamics of
physical quantities at the GDH and at $\spr$ approach {\em universal
curves.} By promptly, we mean the characteristic width of the
ingoing pulse is less than that of the initial GDH (more precisely,
$w/M^\star \lesssim 0.1$). This is most clearly demonstrated in the
behavior of the flux $F^\star$, or equivalently the Bondi mass
$M^\star_{\Bondi}$, measured at $\spr$. An appropriately shifted
affine parameter $y^-_{\rm sh} = y^- + const$ provides an
invariantly defined time coordinate and Fig.~\ref{mass_flux} shows
the universality of evolution of $F^\star$ and $M^\star_{\rm Bondi}$
with respect to it. The shift aligns the $y^-$ coordinates amongst
the solutions, which we are free to do as $y^-$ is only uniquely
defined to within a (physically irrelevant) additive constant.
Finally, we note that this universality is qualitatively different
from the known uniqueness results for solutions of certain
simplified soluble models \cite{rst}. It occurs only if the black
hole is initially macroscopic, formed by a prompt collapse. And in
this case, after the transient phase, the behavior of physical
quantities at $\spr$ does not depend even on the mass.

The overall situation with universality bares some parallels
to the discovery of critical phenomena at the threshold of
gravitational collapse in classical general relativity \cite{mc}
where universal properties were discovered in a system that, at the
time, seemed to have been already explored exhaustively. Of course,
numerical investigations cannot \emph{prove} universality; here we
only studied two families of initial data. However, since these
families, in particular the distribution, are not `special' in any
way, we believe this is strong evidence that universality 
is a feature of the `pure' quantum decay of a GDH, pure in the sense
that the  decay is not contaminated by a continued infall from
$\smr$.

Finally, along the last ray, our simulations showed that curvature
remains finite. Thus, contrary to wide spread belief, based in part
on \cite{hs}, and in contrast to simplified and soluble models,
there is no `thunderbolt singularity' in the metric.
\smallskip

\noindent{\bf{\em \secconclusions. Conclusions.}} In the external
field approximation, the energy flux is initially zero and, after
the transient phase, quickly asymptotes to the Hawking value $F_{\rm
Haw} = \N\hbar\k^2/2 \equiv 0.5$ for the constants used in the
simulations shown here. In the MFA calculation, on the other hand,
at the end of the transient phase the energy flux is \emph{higher}
than this value, keeps monotonically increasing and is about 70\%
greater than $F_{\rm Haw}$ when $M_{\rm Bondi} \sim 2\N M_{\rm Pl}$
(see Fig \ref{mass_flux}). One might first think that the increase
is because, as in 4D, the black hole gets hotter as it evaporates.
This is \emph{not} so: For CGHS black holes, $T_{\rm Haw}\! =
\k\hbar/2\pi$ and $\k$ is an absolute constant. Rather, the
departure from $F_{\rm Haw}\! = 0.5$ shows that, once the back
reaction is included, the flux fails to be thermal at the late stage
of evaporation, \emph{even while the black hole is macroscopic}.
This removes a widely quoted obstacle against the possibility that
the outgoing quantum state is pure in the full theory.

In the classical solution, $\spr$ is \emph{complete} and its causal
past covers only a part of space-time; there is an event horizon.
But $\spr$ is smaller than $\sml$ in a precise sense: $z^-$, the
affine parameter along $\sml$, is finite at the future end of
$\spr$. This is why pure states on $\sml$ of a {\em test} quantum
field $\h{f}_-$ on the classical solution evolve to mixed states on
$\spr$ \cite{atv,vat}, i.e., why the $S$ matrix is non-unitary. In
the MFA, by contrast, our analysis shows that as expected $y^-$ is
\emph{finite} at the last ray on $\spr$. Thus, $\spr$ is incomplete
whence we cannot even ask if the semi-classical space-time admits an
event horizon; \emph{what forms and evaporates is, rather, the GDH.}
However, this incompleteness also opens the possibility that
$\sprb$, the right null infinity of the full quantum space-time, may
be larger than $\spr$ and unitarity may be restored. Indeed, since
there is no thunderbolt, space-time can be continued beyond the last
ray. In the mean field theory the extension is ambiguous. But it is
reasonable to expect that the ambiguities will be removed by full
quantum gravity \cite{swh}. Indeed, since we only have $(0.864/24)
M_{\rm Pl}$ of Bondi mass left over at the last ray \emph{per
evaporation channel} (i.e., per scalar field), it is reasonable to
assume that this remainder will quickly evaporate after the last ray
and $\mbatv$ and $\fatv$ will continue to be zero along the quantum
extension $\sprb$ of $\spr$. Form of $\fatv$ now implies that
$\sprb$ is `as long as' as $\sml$ and hence the $S$-matrix is
unitary: The vacuum state on $\sml$ evolves to a many-particle state
with \emph{finite} norm on $\sprb$ \cite{atv,vat}. Thus unitarity of
the $S$ matrix follows from rather mild assumptions on what
transpires beyond the last ray.

Note, however, this unitarity of the $S$-matrix from $\smr$ to the
extended $\spr$ does \emph{not} imply that all the information in
the infalling matter on $\smr$ is imprinted in the outgoing state on
$\sprb$. Indeed, the outgoing quantum state is completely determined
by the function $y^-(z^-)$ and our universality results imply that,
on $\spr$, this function only depends on $M_{\ADM}$ and not on
further details of the matter profile \cite{apr}. Since only a tiny
fraction of Planck mass is radiated per channel in the portion of
$\sprb$ that is not already in $\spr$, it seems highly unlikely that
the remaining information can be encoded in the functional form of
$y^-(z^-)$ in that portion. Thus, information in the matter profile
on $\smr$ will not all be recovered at $\sprb$ even in the full
quantum theory of the CGHS model. This contradicts a general belief;
indeed, because the importance of $y^-(z^-)$ was not appreciated and
its universality was not even suspected, there have been attempts at
constructing mechanisms for recovery of this information \cite{st}.

To summarize, in 2D there are two distinct issues: i) unitarity
of the S-matrix from $\sml$ to $\sprb$; and ii) recovery of the
infalling information on $\smr$ at $\sprb$. The distinction
arises because right and left pieces of $\mathcal{I}^\pm$ do
not talk to each other. In 4D, by contrast, we only have one
$\mathcal{I}^-$ and only one $\mathcal{I}^+$. Therefore if the
S-matrix from $\mathcal{I}^-$ to $\mathcal{I}^+$ is unitary,
all information in the ingoing state at $\mathcal{I}^-$ is
automatically recovered in the outgoing state at
$\mathcal{I}^+$. To the extent that the CGHS analysis provides
guidance for the 4D case, it suggests that unitarity of the
S-matrix should continue to hold also in 4D \cite{vat}.

\medskip
\centerline{\bf Acknowledgements} \smallskip We would like to thank
Amos Ori and Madhavan Varadarajan for discussions. This work was
supported by the NSF grants PHY-0745779, PHY-0854743, the Eberly
research funds of Penn State, and the Alfred P. Sloan Foundation.


\begin{thebibliography}{99}

\bibitem{cghs} C.~G.~Callan, S.~B.~Giddings, J.~A.~Harvey and
    A.~Strominger,
    Phys. Rev. \textbf{D45}, R1005-R1009 (1992).
\bibitem{reviews} S.~B.~Giddings, {\em arXiv:Hep-th/9412138};
    A.~ Strominger, {\em arXiv:Hep-th/9501071}
\bibitem{hs} S. W. Hawking and J. M. Stewart,
    Nucl.  Phys. \textbf{B 400}, 393-415 (1993)
\bibitem{atv} A. Ashtekar, V. Taveras and M. Varadarajan, Phys. Rev.
    Lett. \textbf{100}, 211302 (2008)
\bibitem{apr} A. Ashtekar, F. Pretorius and F. M. Ramazano\u{g}lu,
    \texttt{arXiv:1011.1024}
\bibitem{pr} F. M. Ramazano\u{g}lu and F. Pretorius,
   {\em arXiv:1009.1440}
\bibitem{vat}M. Varadarajan and A. Ashtekar (pre-print, 2010)
\bibitem{sh}S.~Hayward, Class. Quant. Grav. \textbf{10}, 985-994
    (1993).
\bibitem{st} L.~Susskind and L.~Thorlacius, Nucl. Phys.
    \textbf{B 382}, 123-147 (1992)
\bibitem{dl} D.~A.~Lowe, Phys. Rev. \textbf{D 47}, 2446-2453 (1993)
\bibitem{tu}T.~Tada and S.~Uehara, Phys. Rev. D \textbf{51}
        4259-4264 (1995)
\bibitem{Ori:2010nn}  A.~Ori, Phys.\ Rev.\ D {\bf 82}, 104009 (2010)
\bibitem{ps} T.~Piran and A.~Strominger, Phys. Rev. D\textbf{48},
    4729-4734 (1993)
\bibitem{rst} J.~Rousso, L.~Susskind and L.~Thorlacius, Phys. Rev.
    D\textbf{46}, 3444 (1992)
\bibitem{mc} M.~Choptuik, Phys.\ Rev.\ Lett.\  {\bf 70}, 9 (1993)
\bibitem{swh}S.~W.~Hawking, Commun. Math. Phys. \textbf{43}, 199-220
    (1975)

\end{thebibliography}
\end{document}